\documentclass[twocolumn,showkeywords,preprintnumbers,amsmath,amssymb]{revtex4}

\usepackage{graphicx}
\usepackage{dcolumn}
\usepackage{bm}

\begin{document}

\title{Kant's Copernican Revolution}

\author{Daegene Song}

\affiliation{%
Department of Management Information Systems, Chungbuk National University, Cheongju, Chungbuk 28644, Korea
}%

\date{\today}

\begin{abstract}

A number of philosophers and scientists have discussed the possibility of inseparability between the subject (i.e., the observer) and the object (i.e., the observed universe). In particular, it has recently been proposed that this inseparability may be obtained through the discrete physical universe being filled with the observer's continuous consciousness through quantum evolution with time going backwards. The proposal of a universe view with interwoven matter and mind through cyclical time bears a resemblance to Immanuel Kant's discussion of the Copernican Revolution in philosophy, where the priority shifted from the object to the subject.

\end{abstract}

\maketitle

\section{Introduction}

New scientific theories do not usually appear out of nowhere. First, scientists collect as much evidence as possible and examine the current model. However, the presence of some data that cannot be explained using the current view makes people begin to question the validity of the model. Ordinarily, only a slight modification or extension is proposed, and this process provides the gradual expansion and development of the model. However, some problems with the data may be the consequence of more than a simple misunderstanding or not having enough knowledge about the model; indeed, something more fundamental may be at issue.  

Classical physics, prior to quantum theory, was focused on finding patterns in the physical system. That is, science was trying to find objective truths about the universe in which we live. However, with the development of quantum theory, this approach changed radically (Peres, 1997). Instead of objectivity ruling, quantum theory focused on the way the observer observes the physical system. Indeed, science started to describe the subjective experience rather than the objective reality. Yet classical physics may, too, have possessed a hidden subjective element from the very beginning. However, with great advancements in precision in the twentieth century, people were forced to directly face the subjective limit.

\begin{figure}
\begin{center}
\includegraphics[width=0.4\textwidth]{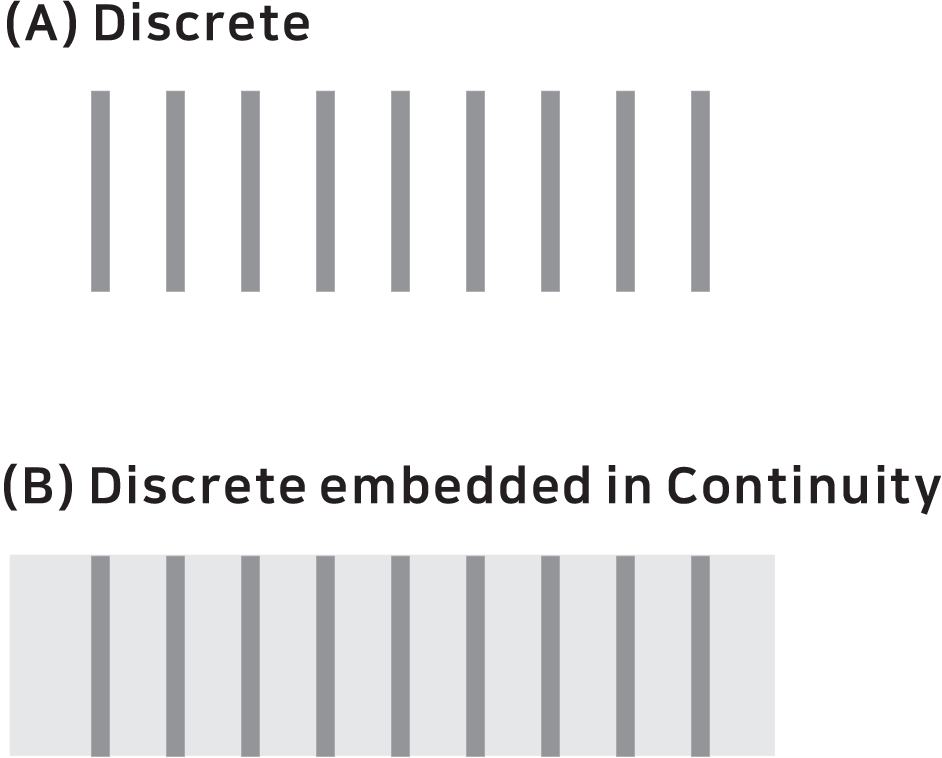}

\end{center}
\caption{ (A) The physical universe may be associated with discreteness. (B) The cyclical-time view of the universe suggests the discrete physical is embedded in the continuous consciousness of the observer.         }

\end{figure}

  Many physicists were stunned by this radical shift, and some people could not accept this subjective scientific theory. In fact, the prominent physicist Albert Einstein was among the scientists who could not accept a subjective quantum theory as the final theory or as a complete scientific law (Einstein et al., 1935). While Einstein was known to be a strong proponent of objective reality, he did mention about the subjectivity that may play a role in science (Einstein, 1932): 
\begin{center}
{\it{Science as something already in existence, already completed, is the most objective, impersonal thing that we humans know. Science as something coming into being, as a goal, is just as subjectively, psychologically conditioned as are all other human endeavors.}}
\end{center}


\section{Subject and Object}
The twentieth-century analytic philosopher Ludwig Wittgenstein proposed the picture theory of meaning, in which he argued that the picture and the reality share the same structure. In particular, Wittgenstein implied that it is language that connects the mental and physical world. For example, Wittgenstein said the following: {\it{naming something is rather like attaching a name tag to a thing}} (Wittgenstein, 1950). In (Song, 2018a), it was argued that language plays a similar role as cyclical time in attaching the physical reality and the conscious understanding of the observed object. For example, when we speak the natural language, which may be equivalent to a discrete physical system (Figure 1 (A)), its continuous meaning is attached as the discrete system being embedded in continuity as in Figure 1 (B). 

Similar to Wittgenstein's argument, the physicist Bohm also discussed the inseparability between individual objects (Bohm, 1980):
\begin{center}
{\it{The notion that all these fragments are separately existent is evidently an illusion … }}  
\end{center}
Moreover, Bohm also emphasized the distinction between the whole and the individual elements:
\begin{center}
{\it{The essential feature in quantum interconnectedness is that the whole universe is enfolded in everything, and that each thing is enfolded in the whole.}} 
\end{center}

On the other hand, Erwin Schr\"{o}dinger discussed the awkwardness in separating objective and subjective realities, which is implied in the Copenhagen interpretation of quantum theory (Schr\"{o}dinger, 1959):
\begin{center}
{\it{The world is given to me only once, not one existing and one perceived. Subject and object are only one. The barrier between them cannot be said to have broken down as a result of recent experience in the physical sciences, for this barrier does not exist.}}
\end{center}

So far, it has generally been the case that science attempts to learn how physical systems work. Moreover, the study of consciousness has also followed this tradition by considering mental processes as a part of natural phenomena. However, this orthodox approach has a fundamental shortcoming because of the self-referential aspect in consciousness (Song, 2007). In fact, understanding the observer's consciousness may be done with respect to the observed physical system rather than as a part of the physical system.


\begin{figure}
\begin{center}
\includegraphics[width=0.4\textwidth]{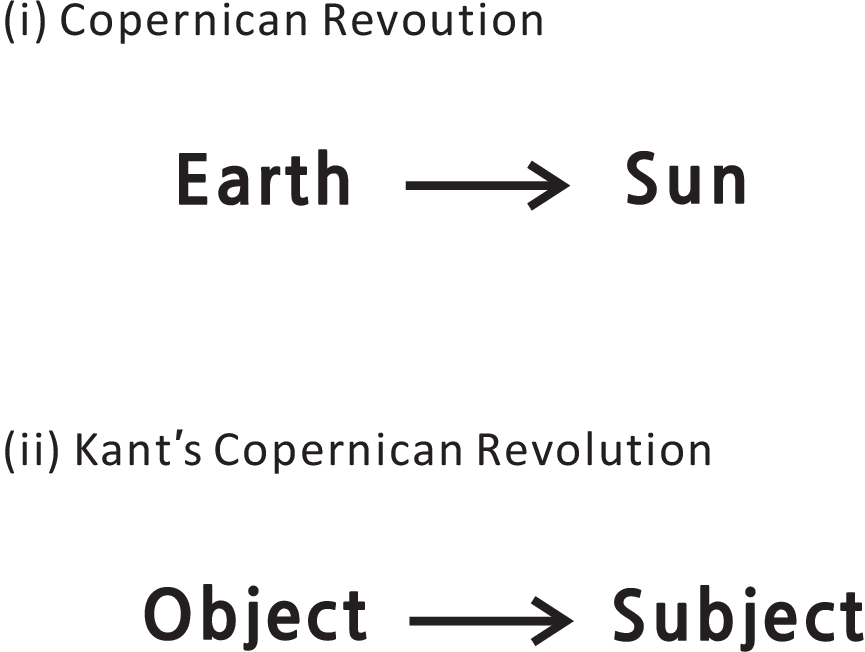}

\end{center}
\caption{ Two types of paradigm shift: (i) A new perspective proposed by Copernicus suggests that it is the sun rather than the Earth that is at the center of our solar system. (ii) In philosophy, Immanuel Kant discussed the emphasis to be placed on the subject rather than the object. This resembles the paradigm shift suggested in Copernican Revolution in science.      }

\end{figure}

\section{Paradigm Shift}
The physicist, philosopher, and historian Thomas Kuhn used the concept of {\it{paradigm}} to explain scientific progress. He proposed three stages within a paradigm: prescience, normal science, and crisis. During the prescience period, there is no dominating platform or paradigm, as multiple candidates exist. Then the community tends to settle into one paradigm, and there is progress made in the normal science stage, in which a substantial amount of progress is made through both theoretical and experimental work confirming the chosen paradigm. Although much evidence supports the chosen paradigm, a different idea then appears that questions the validity of the existing paradigm. When this counter-example accumulates to a certain level, other previously suggested paradigms start to gain more recognition. Then in the final crisis stage, cycles back to the first one, namely, the prescience period.   

Suppose in a closed room, one person is found dead and the following two options are considered. In the first scenario, one person commits suicide, and the second scenario, the other person is murdered. The police start to collect evidence, and a careful analysis of the data is done. 

\begin{figure}
\begin{center}
\includegraphics[width=0.4\textwidth]{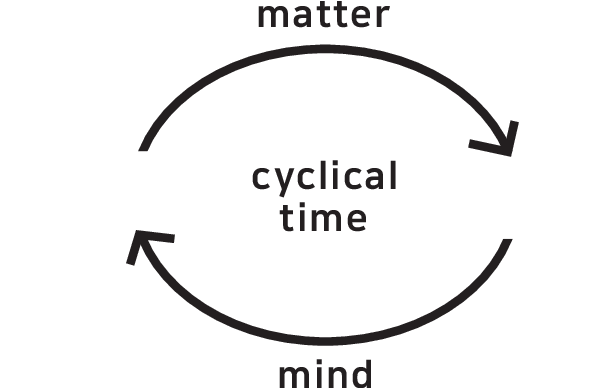}

\end{center}
\caption{ Matter and mind interweaved through cyclical time.      }

\end{figure}

This is the prescience state, as there are competing candidates. The evidence seems to support the first scenario, and the investigators start to think the case as the suicide and gather more evidence and data which continue to support the first scenario. This corresponds to the period of normal science. 

During this period, there is a rapid progress in collecting and analyzing data to support the chosen scenario. However, there starts to appear some evidence or data that contradict the scenario. More evidence appears to yield confusion, and this corresponds to the crisis period. Then critical evidence emerges that eventually confirms that the second murder scenario is correct and that the new paradigm is initiated. One interesting part about this process of paradigm shift is that all the evidence that supported the first scenario, in fact, now supports the second one. Indeed, the evidence was supporting the second paradigm from the very beginning, but it was misinterpreted.  

It is well known that the Copernican Revolution changed the way the universe is viewed. (It should be indicated that as early as the third century BC, Greek astronomer Aristarchus proposed a heliocentric model but did not gain much support.) The philosopher Immanuel Kant discussed a philosophical version of the Copernican Revolution in which the priority shifted from the object to the subject—similar to the view emphasizing the role of the Earth in relation to sun as seen in Figure 2. 

In fact, the radical change of viewing scientific investigation as an interaction between the subject and the object initiated by quantum theory resembles the revolution discussed by Kant. That is, rather than assuming the subject as a part of an object (i.e., the universe), the object of the observed universe and the consciousness of the subject are interwoven (Song, 2017) (Figure 3).


\begin{figure}
\begin{center}
\includegraphics[width=0.4\textwidth]{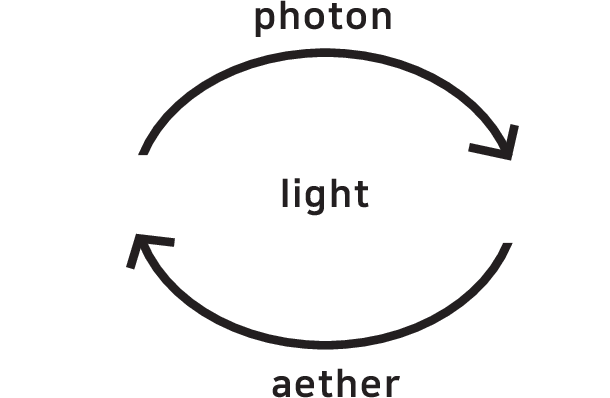}

\end{center}
\caption{ The cyclical-time model hints that photons are physical systems that propagate through the continuous aether of consciousness.    }

\end{figure}

\begin{figure}
\begin{center}
\includegraphics[width=0.4\textwidth]{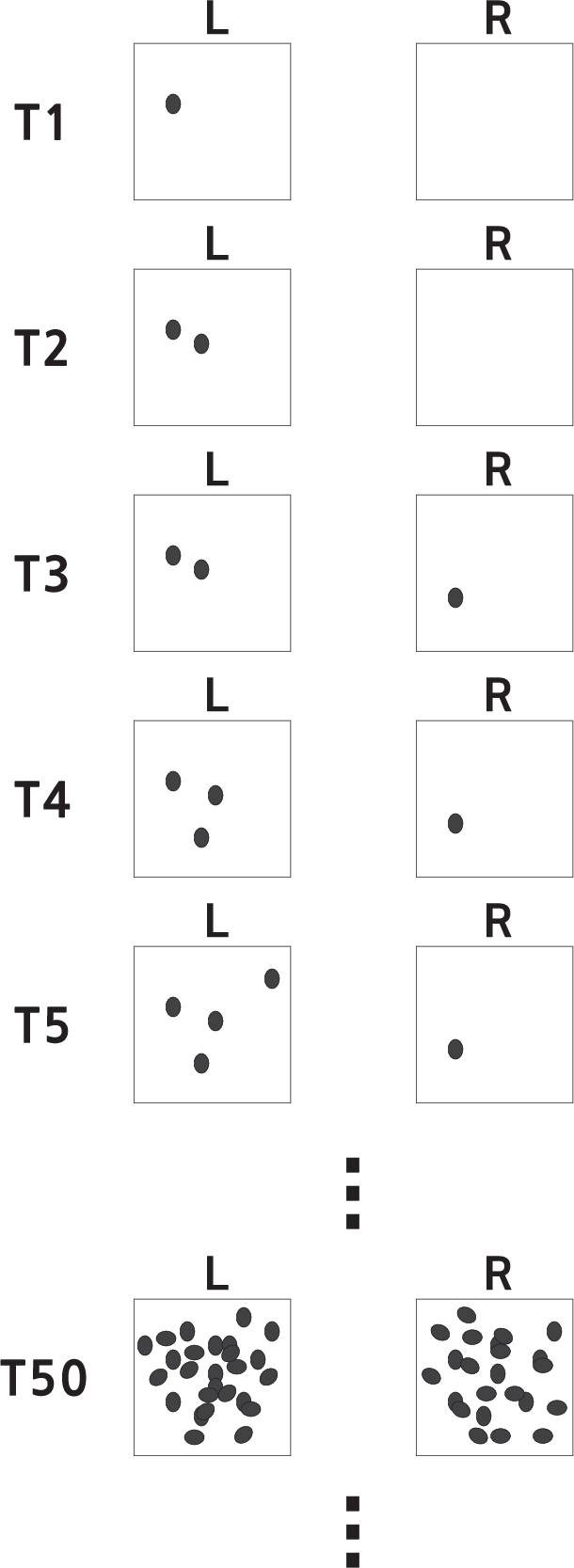}

\end{center}
\caption{ Order and disorder: an example of quantum probability shown through left and right outcomes with equal probability. In the first few trials, the results appear to be random; however, as more outcomes appear, the probabilistic order of 50-50 appear to rise.       }

\end{figure}

The ancient Greek philosopher Aristotle discussed the concept of the aether as one of the fundamental ingredients of the universe. Einstein also mentioned the possibility of the existence of aether: {\it{According to the general theory of relativity space without Aether is unthinkable …}} (Einstein, 1983). Paul Dirac also discussed that quantum vacuum may be associated with aether: {\it{… we are rather forced to have an Aether}} (Dirac, 1951). When asked about the resolution of the EPR paradox (Einstein et al., 1935), John Bell argued that it may be enlightening to go back to the pre-Einstein era to examine relativity and to reconsider the possibility of aether. Moreover, Bell also proposed that the introduction of aether may help to resolve the problem of nonlocality as shown in the EPR paradox. 

As shown in Figure 4, the new subject model (Song, 2017) suggests the puzzling aspect of aether as consciousness experience of the subject in regards to the photon object. For instance, the probabilistic nature of the quantum theory existing in the following state,
\begin{equation}
|\psi\rangle =\frac{1}{\sqrt{2}} \left( |L\rangle + |R\rangle \right) 
\end{equation}
may be understood as randomness and order progressing simultaneously. Thus, the physical process generates disorder, yet the conscious awareness of the process provides order (Figure 5).

Moreover, the philosopher Thales, who predicted the eclipse in 585 BC, contemplated what things are composed of. In fact, his proposal that water is the origin of matter is similar to the model of the physical universe filled with the continuous negative sea of consciousness of the subject.


\section{Time}
Existentialism is a school of thought in philosophy that developed in the twentieth century. The proposed cyclical-time universe and consciousness in (Song, 2017) bears resemblance to a number of ideas in existentialism. The founder of the movement, S\o ren Kierkegaard, contemplated various concepts associated with subjectivity in obtained knowledge. Then Edmund Husserl proposed the method of phenomenology, which studies philosophical questions using the detailed analysis of phenomenon and the conscious observer. In particular, Husserl argued that the conscious observer and the objects being observed may not be separable. Similarly, the cyclical-time universe model proposes that the conscious observer and the physical object being observed are linked together through the cyclical-time process. 

Another existentialism philosopher, Heidegger, argued that the linear line of past, present, and future events is simply wrong. That is, to consider past as {\it{no-longer-now}} and future as {\it{not-yet-now}} may not be valid (Heidegger, 1962) (Figure 6). Instead, he discussed that the present, past, and future ought to be considered as an interwoven unity. Similar to the interdependence of space-time and matter in Einstein's theory of gravitation, Heidegger proposed the interconnectivity between existence and time such that each determines the other. Henri Bergson also mentioned the continuity of time using the notion of duration to distinguish the outer physical world from the inner conscious world (Bergson, 1946).

\begin{figure}
\begin{center}
\includegraphics[width=0.4\textwidth]{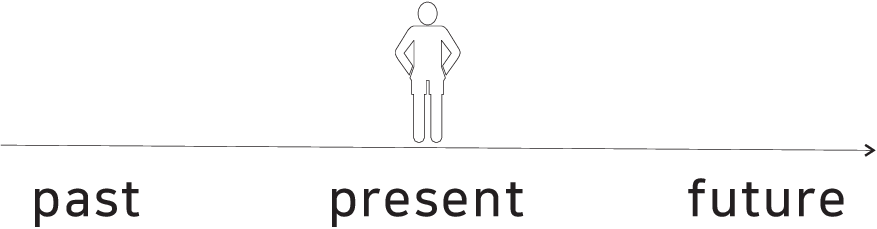}

\end{center}
\caption{ The current time model employed in science is a linear one. As suggested by Heidegger, treating the past as {\it{no-longer-now}} and the future as {\it{not-yet-now}} may not provide a correct picture of time.         }

\end{figure}

\section{Remarks}
In the sixteenth century, Copernicus proposed a concept radically changed people's view of the universe. Ever since quantum theory was introduced about a century ago, there has been confusion on its subjective nature involving probability and measurement. However, as many philosophers and notable physicists, such as John Wheeler (Wheeler, 1990) and David Bohm (Bohm, 1980), have implied, quantum theory initiated a new paradigm of emphasizing the subject, as opposed to the previous view of the subject as only a part of the object. 

Although there has been great improvement in machines acting and thinking like humans, another interesting aspect has been detected as well. That is, even though machines can do tasks that humans have great difficulties, such as moving heavy weights and calculating large numbers, the tasks that humans perform rather easily are often difficult for machines to imitate. For instance, while robots are built to perform delicate tasks, they have difficulty doing a casual walk, which many humans can easily do. Therefore, this trend seems to imply that something very easy, even the easiest task for humans, may be impossible for machines. This distinguishability between machine and humans (Song, 2018b), in fact, provides a new perspective on our universe, which is similar to Kant's discussion.



\begin{thebibliography}{}

\bibitem{1} Bergson H. The creative mind: an introduction to metaphysics. Kensington, New York, 1946.
\bibitem{2} Bohm D. Wholeness and the implicate order. London: Routledge \& Kegan Paul, 1980. 
\bibitem{3} Dirac PAM. Is there an Æther? Nature 1951; 168: 906-907.
\bibitem{4} Einstein A. Address to Students of UCLA, 1932.
\bibitem{5} Einstein A. Ether and the theory of relativity, Address delivered in the university of Leyden, May 5th, 1920, in Sidelights on relativity, (Dover, New York, 1983) 
\bibitem{6} Einstein A, Podolsky B, Rosen N. Can quantum-mechanical description of physical reality be considered complete? Phys Rev 1935; 47: 777-780. 
\bibitem{7} Heidegger M. Being and time. New York: Harper and Row, 1962. Mathematica und verwandter Systeme I. Monatshefte fur Mathematik Und Physik 1931; 38(1):173-98.
\bibitem{8} Peres A. Quantum theory: concepts and methods, Kluwer, Dordrecht, 1997.
\bibitem{9} Shr\"{o}dinger E, Mind and matter, Cambridge University Press, 1959.
\bibitem{10} Song D. Non-computability of consciousness. NeuroQuant 2007; 5: 382-391. arXiv:0705.1617 [quant-ph].
\bibitem{11} Song D. Decision-Making process and information. NeuroQuant 2017; 15: 31-36. 	arXiv:1701.08641 [physics.gen-ph].
\bibitem{12} Song D. Encryption and information network. NeuroQuant 2018a; 16: 1-6.
\bibitem{13} Song D. Machine vs human: similarities and differences. NeuroQuant 2018b; 16: 87-91.
\bibitem{14} Wheeler JA. Information, physics, quantum: the search for links. In Zurek ed., Complexity, entropy, and the physics of information. Redwood City, California: Addison-Wesley, 1990.
\bibitem{15} Wittgenstein L. Tractatus logico-philosophicus, International Library of Psychology, Philosophy, and Scientific Method. Trans. CK Ogden, 8th impr. London, Routledge \& Kegan Paul, 1922.
\bibitem{16} Wittgenstein L. Philosophical investigations. New York: Macmillan, 1953.





\end{thebibliography}
\end{document}